# Development of precession Lorentz transmission electron microscopy

Shunsuke Hayashi[A], Dongxue Han[A,B],

Hidenori Tsuji[B], Kyoko Ishizaka[A,B], and Asuka Nakamura[A,B,†]

† corresponding author. Email: asuka.nakamura@riken.jp

[A]*Quantum-Phase Electronics Center and Department of Applied Physics, The University of Tokyo,*

*Bunkyo, Tokyo 113-8656, Japan*

[B]*RIKEN Center for Emergent Matter Science (CEMS), Wako, Saitama, 351-0198, Japan*

**Abstract**

Lorentz transmission electron microscopy (LTEM) is a powerful tool for high-resolution imaging of magnetic textures, including their dynamics under external stimuli and ultrafast nonequilibrium conditions. However, magnetic imaging is often hindered by non-magnetic diffraction contrast arising from inhomogeneous sample deformation or a non-parallel electron beam. In this study, we develop a precession LTEM system that can suppress diffraction contrast by changing the incident angle of the electron beam relative to the sample in a precessional manner. By comparing LTEM images acquired at different precession angles ($\theta$), we show that diffraction contrast is significantly reduced with increasing $\theta$. However, large $\theta$ values lead to an undesired broadening of the magnetic contrast, highlighting the importance of optimizing $\theta$. Furthermore, defocus-dependent measurements reveal that magnetic contrast is particularly improved at small defocus values, suggesting that precession LTEM can achieve higher spatial resolution. These findings demonstrate the potential of precession LTEM as a powerful technique for studying magnetic dynamics.

Keywords: precession Lorentz transmission electron microscopy, magnetic imaging, diffraction contrast

# 1. Introduction

Lorentz transmission electron microscopy (LTEM) is a powerful technique for investigating magnetic textures with high spatial resolution. Electrons passing through a ferromagnetic material are deflected by the Lorentz force, and the magnetic domain wall can be visualized as dark/bright contrast in the defocused images [Fig. 1(a)], as revealed by the pioneering work by Hale *et al*. [1]. In addition to the conventional ferromagnetic domains, LTEM has been used to observe various types of spin textures [2], such as magnetic stripes and skyrmions [3,4]. The development of in-situ observation techniques under various external stimuli, such as electric current [5,6] and strain [7–9], also plays a crucial role in understanding magnetic dynamics. This is evidenced by diverging recent studies such as the electric-current-induced motion of magnetic domain walls and topological spin textures [10–12], strain-induced skyrmion deformation [13], and heat-current-driven skyrmion motion [14]. Furthermore, the recently developed ultrafast transmission electron microscopy, utilizing the pump-probe method, has enabled the observation of nonequilibrium dynamics of spin textures on ps-ns timescales [15–19].

However, diffraction contrast (so-called bend contour), which arises from inhomogeneous electron diffraction conditions due to sample deformation or a non-parallel electron beam [Fig. 1(b)], hinders the observation of magnetic dynamics. Such diffraction contrast interferes with magnetic contrast, reducing the accuracy of magnetic structure measurements. Conventionally, this problem has been avoided by adjusting the sample tilt to move the diffraction contrast outside the region of interest. However, this is not enough in several situations, such as in-situ measurements and time-resolved TEM measurements, where the magnetic/diffraction contrast can be significantly changed by the external stimuli. Therefore, a technique for suppressing diffraction contrast is highly required to investigate spin textures and their dynamics.

In this study, we develop a precession LTEM method that can realize the suppression of diffraction contrast. In this method, the incident angle of the electron beam is tilted by $\theta$ from the optical axis, and the azimuthal angle $\phi$ is precessionally varied around the optical axis during data acquisition [Fig. 1(c)], which results in a significant alteration of diffraction conditions. By averaging LTEM images under different diffraction conditions, it is expected that diffraction contrast can be strongly suppressed while the magnetic contrast is not significantly changed. The precession and similar tilt-scan techniques have been developed in the electron diffraction community [20–22] and recently applied to imaging techniques such as differential phase contrast (DPC) [23–25] and bright-field imaging [26]. In this study, we apply the precession technique to LTEM and evaluate its performance in suppressing diffraction contrast. First, we investigate the effect of precession angle $\theta$ on diffraction and magnetic contrast by comparing the $\theta$-dependent precession LTEM images. Second, we compare the defocus ($\Delta f$)-dependences of the conventional and precession LTEM images. Through this comparison, we demonstrate the improvement of magnetic imaging, which is particularly advantageous for observations in small-defocus regions. These results demonstrate the advantage of precession LTEM, which will play a crucial role in future investigations of spin texture dynamics.

## 2. Development of precession LTEM

### 2.1 Setup of precession LTEM

Figure 1(c) shows a schematic of the precession LTEM measurement system developed in this study. In precession LTEM, the electron beam is tilted off the optical axis by a precession angle $\theta$ and precessed with an angle $\phi$ around the optical axis. We control $\phi$ by a software and record the $\phi$-depenent LTEM images $I(\phi)$. Under the defocused condition, changing $\phi$ causes a shift of the electron beam relative to the camera (screen), as shown in Fig. 2(a). Therefore, it is necessary to correct these electron-beam shifts during the measurement so that the electron beam stays inside the camera view. To address this, we adjust the electron deflector in the TEM for each $\phi$ to correct the image position on the camera, by analyzing the recorded LTEM image. Figure 2(b) illustrates the method for extracting the amount of the electron-beam shift. Each LTEM image (left) is differentiated using the Prewitt filter (middle) and then segmented with simple thresholds to extract only the sample edges (right). We compare this image with a reference image measured in advance, by using a cross-correlation algorithm. In this way, we determine the amount of the shift.

Figure 2(c) shows the measurement procedure. First, we acquire $I(\phi_0)$ at the initial angle $\phi_0$. After rotating the beam by $\Delta\phi$ to the next angle $\phi_1$, we acquire $I(\phi_1)$. Because $\Delta\phi$ is sufficiently small (1°), the electron-beam shift $\Delta r$ between $I(\phi_0)$ and $I(\phi_1)$ is less than few pixels, ensuring that the region of interest remains within the field of view of the camera. After recording $I(\phi_1)$, we calculate the amount of the beam shift $\Delta r$ based on the method described above and adjust the beam shift using the deflector so that the edge of the sample aligns with the same position as in the reference image $I(\phi_0)$. By repeating this process, we can obtain the $\phi$-dependent LTEM images. Since these LTEM images are obtained before the electron-beam shift is fully corrected, each $I(\phi)$ obtained in this procedure is slightly shifted compared to $I(\phi_0)$. Therefore, we correct these slight shifts before integrating $I(\phi)$ over $\phi$ using the same calculation method described above. This measurement procedure is advantageous for reducing acquisition time by half compared to a method in which each image is captured after the electron-beam shift is fully corrected. Since each $I(\phi)$ has a different diffraction contrast, integrating $I(\phi)$ should significantly reduce the diffraction contrast without substantially affecting the magnetic contrast.

**2.2 Experimental setup**

The detailed setup of our TEM is described elsewhere [22]. As a pixelated detector, we use the MerlinEM (Quantum Detectors) of 256 × 256 pixels. As a cathode, we use 50 μm carbon-coated $LaB_6$ (Applied Physics Technologies). We use a 150 μm condenser aperture. We use a 100 nm thin plate of $(Fe_{0.63}Ni_{0.30}Pd_{0.07})_3P$ (FNPP) [27] prepared by the focused ion beam method. The (1 -1 0) oriented ferromagnetic thin plate is in-plane magnetized in the absence of external magnetic field. We set the precession angle $\theta$ between 0.01 and 0.40°. The number of divisions for $\phi$ is 360 ($\Delta\phi = 1°$). As a result, $I(\phi)$ is obtained as a 256 × 256 × 360 three-dimensional datasets. For each $\phi$, we acquire an individual LTEM image with an integration time of 1 s. After correcting for the image shift, we integrate $I(\phi)$ along $\phi$ and obtain the precession LTEM image. Therefore, the recorded time per one precession LTEM image is 360 s. We also obtain a conventional LTEM image without precession with an integration time of 360 s to compare it with the precession LTEM result.

## 3. Results

### 3.1 Effect of precession angle on diffraction and magnetic contrast

To begin with, we investigate the effect of precession angle $\theta$ on diffraction and magnetic contrast. Figure 3(a) shows the conventional ($\theta = 0$) and $\theta$-dependent precession LTEM images of a ferromagnetic domain wall in FNPP. The vertical line with dark contrast observed in the image corresponds to the ferromagnetic domain wall in this material. In the surrounding region, additional non-uniform dark contrasts are observed, which can be interpreted as diffraction contrast. As compared with the conventional LTEM results, diffraction contrast is strongly suppressed in the precession LTEM images, particularly at $\theta > 0.10°$. The line profiles along arrow #1 in Fig. 3(a) show a systematic decrease in the diffraction contrast [Fig. 3(b)]. In Fig. 3(c), we further investigate the effect of $\theta$ on magnetic contrast from the line profiles along arrow #2, which shows the increase in the width of the magnetic contrast at the large $\theta$ region. These results indicate that diffraction contrast is successfully suppressed by precession LTEM while there is an undesirable broadening of magnetic contrast at large $\theta$.

To understand the effect of $\theta$ on diffraction contrast, we quantitatively evaluate its change. The suppression of diffraction contrast is evaluated using the noise power spectrum of the line profiles. From each LTEM image in Fig. 3(a), we obtained $P_y(k_y)$ by calculating the noise power spectrum along the $y$ direction as:

$$P_y(k_y) = \int dx \, \mathcal{F}_y[I(x,y)]^2$$

where $I(x,y)$ is an experimentally obtained intensity at the position $(x,y)$, $\mathcal{F}_y$ is Fourier transform along the $y$ direction. The integration range in the $x$ direction is over the entire image. For the Fourier transformation, we applied the fast Fourier transform algorithm with a Blackman window implemented in *lys* [28]. For an ideally uniform LTEM image without diffraction contrast, $P(k) = \delta(k)$ should be satisfied. In the present case, $P_y(k_y \neq 0)$ predominantly reflects the diffraction contrast since the magnetic contrast is not observed along the $y$ direction. Accordingly, the decrease in $P_y(k_y)$ seen in Fig. 3(d) can be interpreted as $\theta$-dependent suppression in diffraction contrast. Figure 3(e) further shows that $P_y(k_y = 3.0 \text{ μm}^{-1})$ is more rapidly suppressed as $\theta$ increases, compared to $P_y(k_y = 1.0 \text{ μm}^{-1})$. Considering that smaller (larger) $k_y$ corresponds to diffraction contrast with a larger (smaller) spatial scale, this result indicates that diffraction contrast with a smaller spatial scale is more rapidly suppressed as $\theta$ increases. In the large $\theta$ (> 0.20°) region, $P_y(k_y)$ is nearly constant for all $k_y$, indicating that the suppression is saturated.

We further quantitatively analyze the broadening of the magnetic contrast in the precession LTEM images as a function of $\theta$, by fitting each line profile with a Gaussian function and a linear background [solid curves in Fig. 3(c)]. Figure 3(f) shows the Gaussian width σ obtained from the fitting, which is significantly increased at $\theta > 0.10°$. This indicates that the magnetic contrast in precession LTEM image is broadened at larger $\theta$. Such broadening should be explained by the variation of the spherical aberration depending on $\phi$ during precession and may be improved by introducing $\phi$-dependent aberration correction. Present results demonstrate the influence of $\theta$ on diffraction and magnetic contrast in precession LTEM: The suppression of diffraction contrast requires larger $\theta$ while magnetic contrast is simultaneously broadened. In the present case, we found that the magnetic contrast is not significantly broadened at $\theta = 0.10°$ while the diffraction contrast is strongly suppressed. Therefore, we use $\theta = 0.10°$ in the next section.

### 3.2 Defocus-dependent conventional and precession LTEM images

To understand the advantages of precession LTEM, we focus on the defocus ($\Delta f$) dependence of both conventional and precession LTEM results at several diffraction conditions. In $\Delta f$-dependent conventional LTEM images [Fig. 4(a)], ferromagnetic domain walls appear as dark contrast in the overfocus images ($\Delta f > 0$), which are inverted in the underfocus images ($\Delta f < 0$). Increasing $\Delta f$ enhances the magnetic contrast without significantly affecting the diffraction contrast, as expected. In the $\Delta f$-dependent precession LTEM images with $\theta = 0.10°$ [Fig. 4(b)], diffraction contrast is strongly suppressed as compared to the conventional LTEM images. Since the diffraction contrast in the conventional LTEM images in Fig. 4(a) is significantly dependent on the sample positions, we can investigate the influence of diffraction contrast by analyzing the position-dependent line profiles. When we focus on the region where the diffraction contrast is relatively weak (line profile along #1), the magnetic contrast is more pronounced in the precession LTEM result as compared with that of the $\theta = 0$ case at the same $\Delta f$ [Fig. 4(c,d)]. Such an improvement of magnetic contrast is much more significant when we focus on the line profile along #2, where diffraction contrast is strong: while we cannot distinguish the magnetic contrast even at high $\Delta f$ in the conventional LTEM images [Fig. 4(e)], it is clearly observed in the precession LTEM images [Fig. 4(f)]. These results demonstrate that precession LTEM can play an important role, especially when the diffraction contrast is comparable to the magnetic contrast.

We further quantitatively investigate the effect of precession LTEM in the region where the diffraction contrast is relatively weak. Figure 5(a,b) show the normalized power spectrum along the $x$ direction [$\widetilde{P}_x(k_x) = P_x(k_x)/P_x(k_x = 0)$] of the conventional and precession LTEM images in the red rectangles. Since there is a vertical (//$y$) line-shaped magnetic contrast in this region, both magnetic and diffraction contrast contribute to $\widetilde{P}_x(k_x)$. In all $k_x$ ranges, $\widetilde{P}_x(k_x)$ increases monotonically with $\Delta f$, which reflects the enhancement of magnetic contrast. On the other hand, Fig. 5(c,d) show $\widetilde{P}_y(k_y)$ in the same region, which is composed of only diffraction contrast. $\widetilde{P}_y(k_y)$ of precession LTEM [Fig. 5(d)] is significantly smaller than that of conventional LTEM [Fig. 5(c)], reflecting the strong suppression of diffraction contrast. To quantitatively compare the $\Delta f$-dependence of magnetic and diffraction contrast, we plotted $C_{M+D} = \int dk_x\, \widetilde{P}_x(k_x)$ and $C_D = \int dk_y\, \widetilde{P}_y(k_y)$ in Fig. 5(e,f) with the integration range between 3 μm⁻¹ and 10 μm⁻¹. Here, $C_{M+D}$ reflects the combined contribution of magnetic and diffraction contrast, while $C_D$ represents only diffraction contrast, with no contribution from the magnetic contrast. Although $C_{M+D}$ is very similar between conventional and precession LTEM, the reduction in $C_D$ relative to $C_{M+D}$ is more pronounced in precession LTEM. Consequently, compared to conventional LTEM, precession LTEM allows us to observe magnetic contrast at smaller $\Delta f$. In other words, we can obtain a similar $C_{M+D}/C_D$ with smaller $\Delta f$. Since the spatial resolution of LTEM is improved at smaller $\Delta f$, this result means we can achieve higher spatial resolution with precession LTEM.

## 4. Discussion

Finally, we discuss the relationship between precession LTEM and other existing methods in the TEM field. In previous studies, precession and tilt-scan DPC, which can also suppress the diffraction contrast in magnetic imaging, have been developed [29–31]. Compared with these methods, precession LTEM is advantageous because it can be achieved without additional hardware components. In addition, the application of precession LTEM for in-situ and time-resolved measurements under external stimuli is a promising pathway for the investigation of spin texture dynamics. For example, in time-resolved pump-probe measurements using ultrafast TEM [18], the time-dependent changes in diffraction contrast prevent us from obtaining clear magnetic dynamics. A combination of precession LTEM and such time-resolved measurements should play an important role in the investigation of magnetic dynamics in the future.

## 5. Conclusion

In conclusion, we developed a precession LTEM technique and evaluated its performance in suppressing diffraction contrast. We obtained $\phi$-dependent LTEM images, from which we can reconstruct the precession LTEM image with precession angle $\theta$. We evaluated the suppression of diffraction contrast as a function of $\theta$, showing saturation at larger precession angles. Additionally, the magnetic contrast in the precession LTEM images was undesirably broadened at larger $\theta$. It is essential to optimize the precession angle when performing precession LTEM. Further investigation of the defocus ($\Delta f$) dependence demonstrated the improvement in magnetic imaging. Such a method is advantageous in the small defocus region and when the diffraction contrast is comparable to the magnetic contrast. These results indicate the potential of precession LTEM, which will play an important role in the future investigation of spin texture dynamics.

**Note**

The authors declare no competing interest.

**Acknowledgments**

We thank K. Karube, Y. Taguchi, and Y. Tokura for bulk sample preparation. This work was partially supported by Grant-in-Aid for Scientific Research (KAKENHI) (Grant No. 25K00057), JST, PRESTO Grant No. JPMJPR24JA.

**CRediT authorship contribution statement**

**S. Hayashi:** Data curation, formal analysis, investigation, methodology, validation, visualization, writing-original draft. **D. Han:** Formal analysis, investigation, resource, validation, writing-review & editing. **H. Tsuji:** Data curation, software, validation. **K. Ishizaka:** Funding acquisition, supervision, writing-review & editing. **A. Nakamura:** Conceptualization, data curation, funding acquisition, investigation, methodology, project administration, software, supervision, writing-review & editing.

**Figures and Captions**

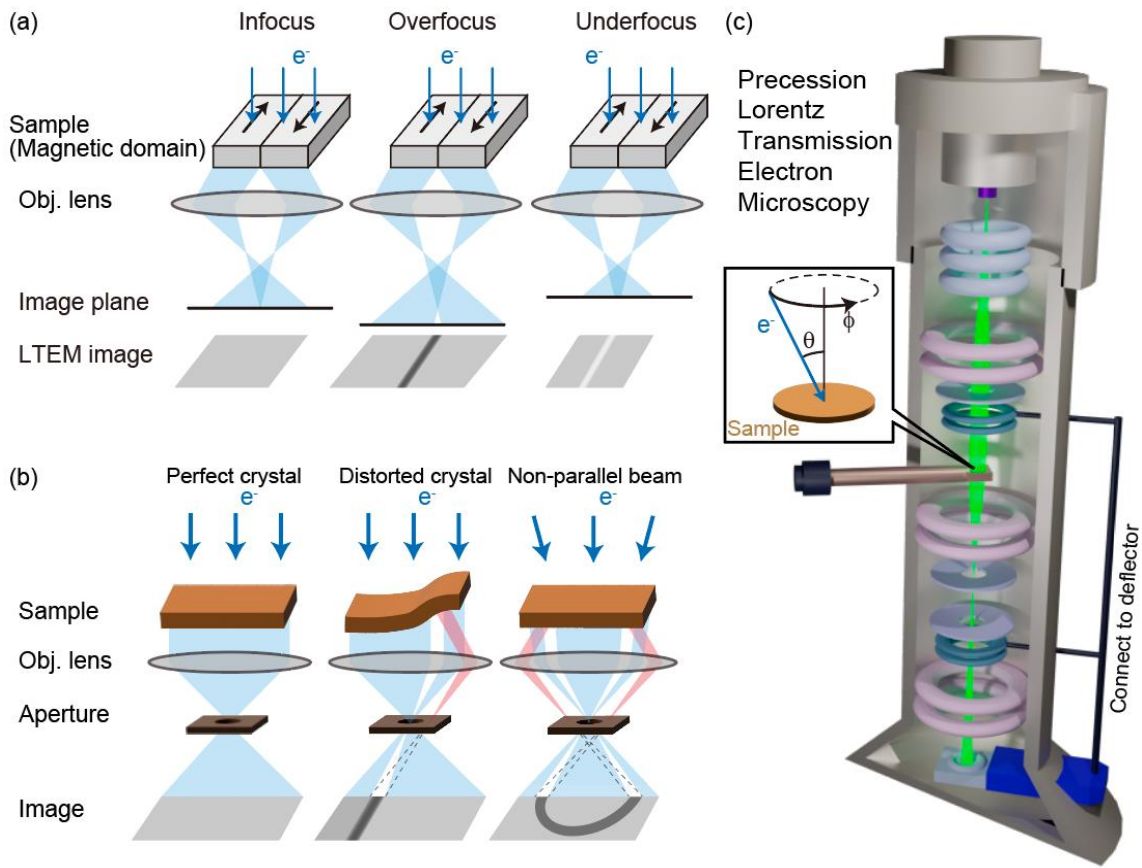

Fig. 1. **Experimental setup.** (a) Visualization of magnetic domain walls in Lorentz TEM under different defocus conditions (infocus, overfocus, and underfocus). The electrons deflected by the Lorentz force form dark/bright contrast at the domain wall position. (b) Schematic of diffraction contrast formation. The diffraction of electrons caused by sample deformation and a non-parallel electron beam condition results in the diffraction contrast. (c) Experimental setup for the precession LTEM. The incident angle of the electron is changed precessionally.

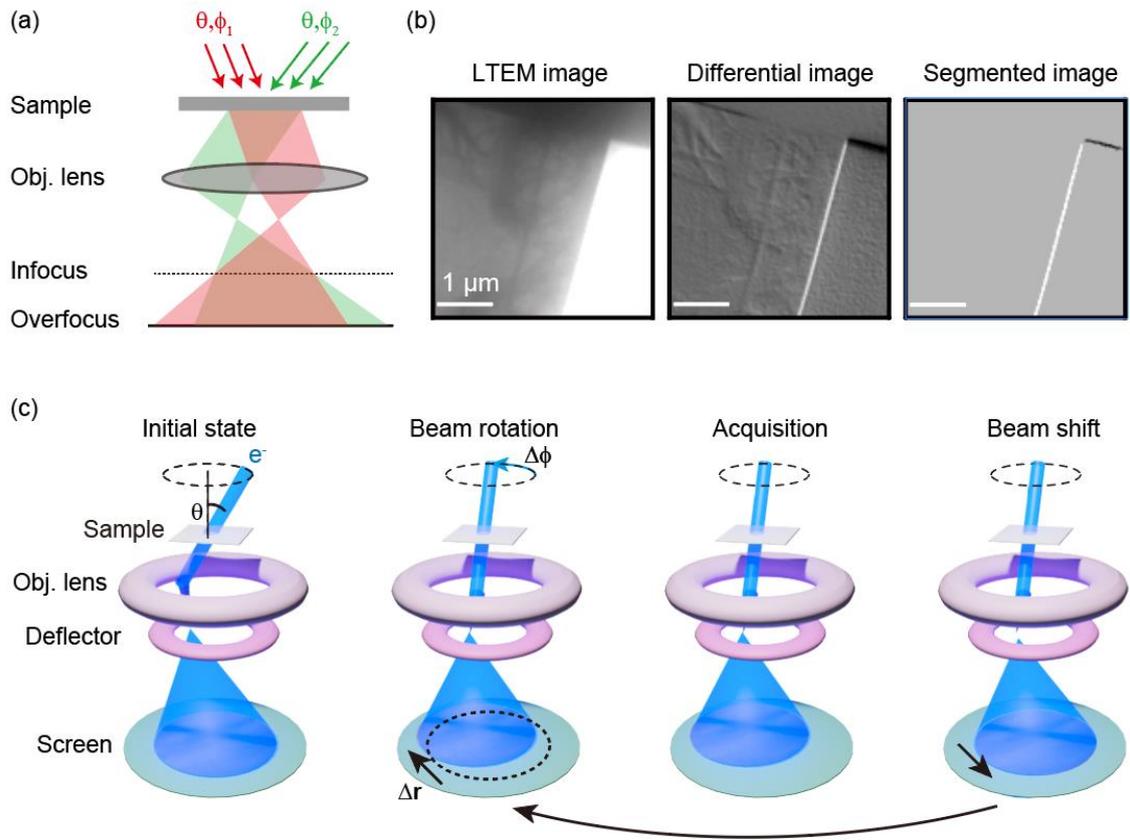

Fig. 2. **Procedure for acquiring $\phi$-dependent LTEM images**. (a) $\phi$-dependent shift of the electron beam with respect to the camera (screen) under defocused conditions. The red and green regions represent the trajectory of the electrons with different $\phi$. (b) Steps for calculating the amount of the beam shift. To extract only the sample edges, each LTEM image (left) is differentiated using the Prewitt filter (middle) and then segmented using thresholding (right). (c) Measurement procedure for acquiring $\phi$-dependent LTEM images. Beam shifts caused by the beam rotation are corrected after image acquisition.

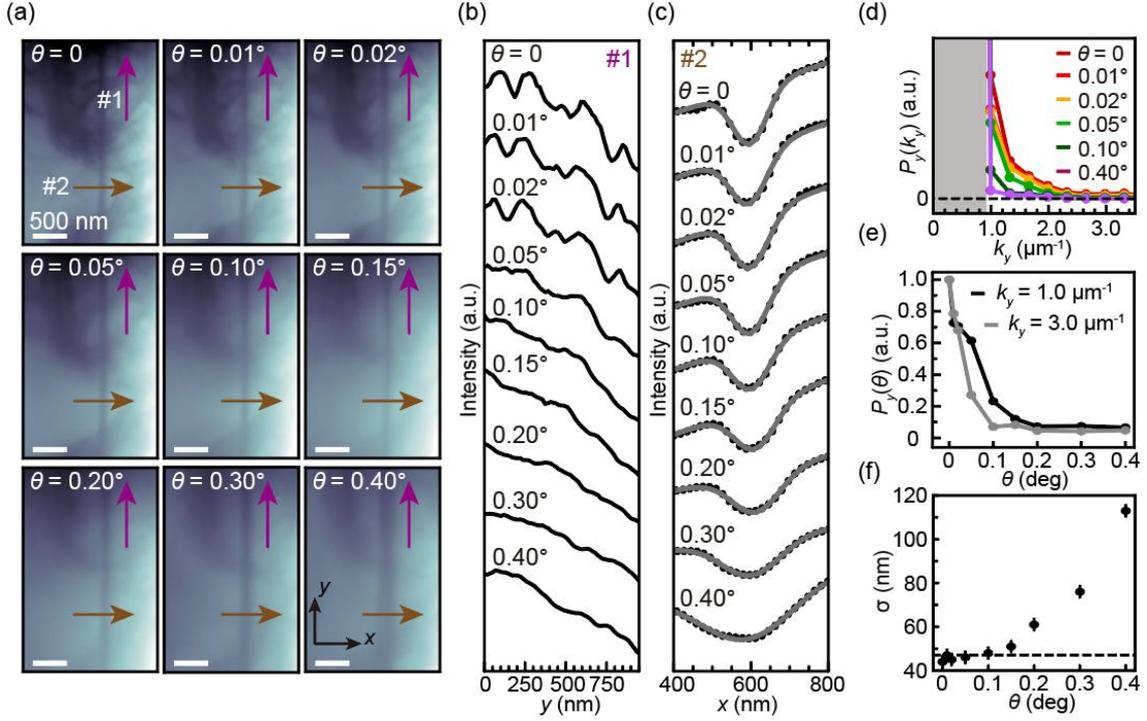

Fig. 3. **Effect of $\theta$ on diffraction and magnetic contrast**. (a) Conventional and $\theta$-dependent precession LTEM images. The vertical-line-shape dark contrast observed in the image corresponds to the ferromagnetic domain wall. The additional nonuniform dark contrasts in the surrounding region are interpreted as diffraction contrast. (b) Line profile along the arrow #1 in a. (c) Line profile along the arrow #2 in a. The gray curve shows the fitting result using a Gaussian and linear background. (d) Noise power spectrum $P_y(k_y)$ of the conventional ($\theta = 0$) and $\theta$-dependent precession LTEM images. $P_y(k_y)$ is normalized by $P_y(k_y = 0)$ for each $\theta$. The gray-shaded region at low $k_y$ ($< 1.0$ μm$^{-1}$) is not used for analysis because it is significantly affected by a peak at $k_y = 0$. (e) $\theta$-dependent $P_y(k_y = 1.0$ μm$^{-1})$ and $P_y(k_y = 3.0$ μm$^{-1})$, normalized by $P_y(\theta = 0)$. (f) $\theta$-dependent Gaussian width σ obtained from the fitting in c.

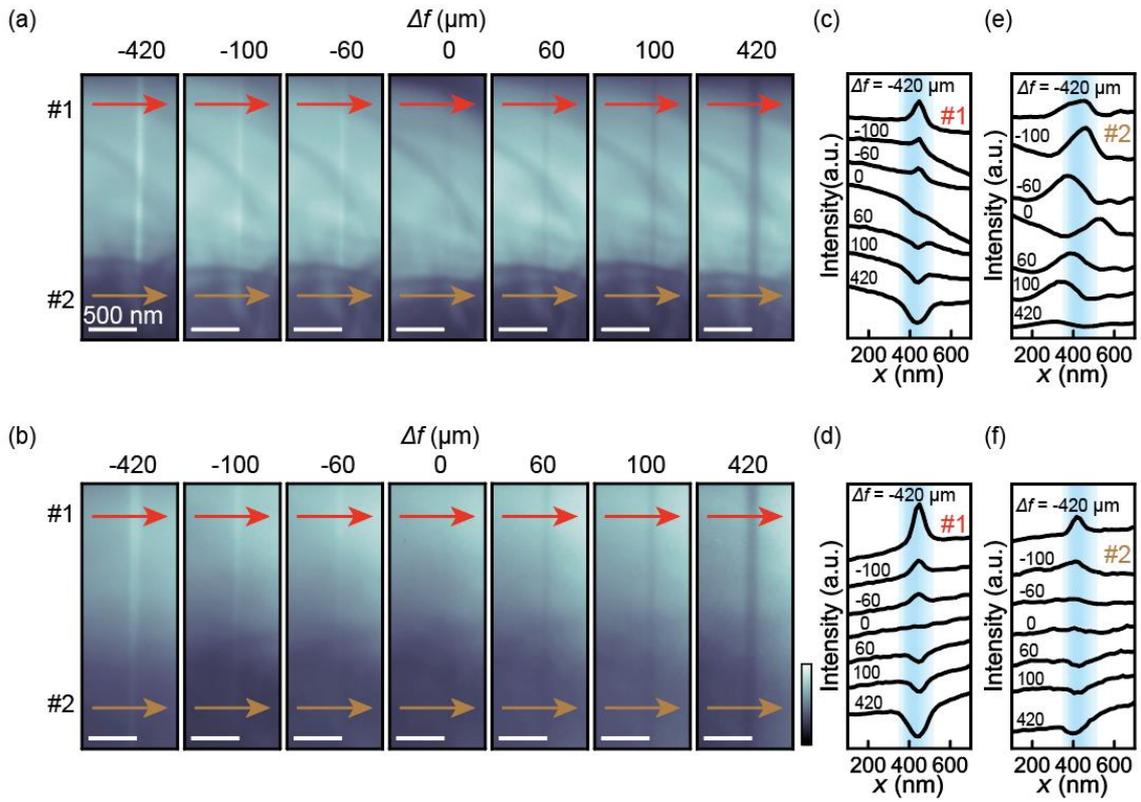

Fig. 4. *Δf* -**dependent conventional and precession LTEM images.** (a,b) *Δf* -dependent conventional and precession ($\theta = 0.10°$) LTEM images. (c,d) Line profiles along the arrow #1 in a and b. (e,f) Line profiles along the arrow #2 in a and b. The blue region indicate the positions of the domain wall.

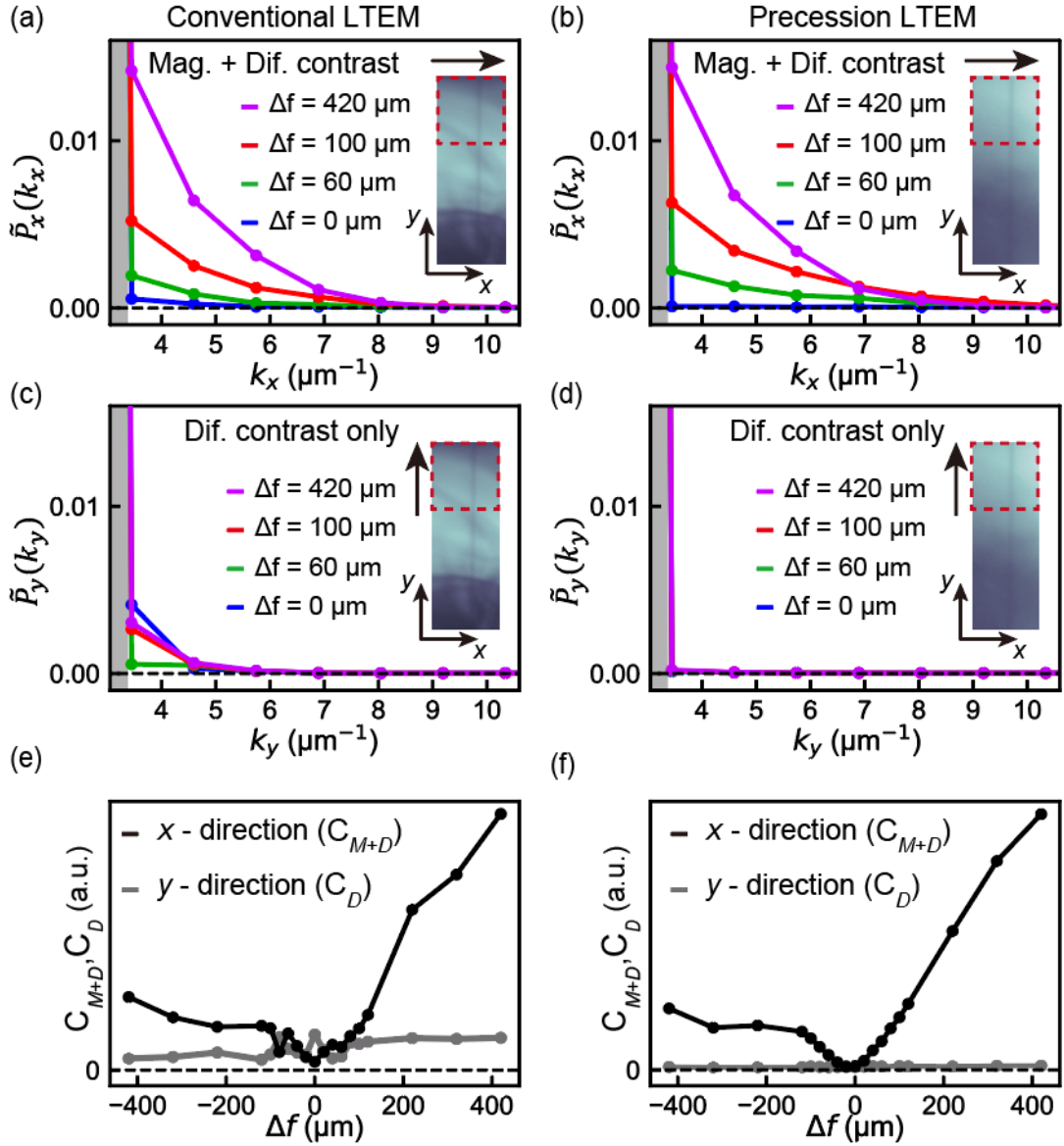

Fig. 5. *Δf*-dependence of magnetic and diffraction contrast. (a,b) Normalized power spectrum along the $x$ direction $[\widetilde{P}_x(k_x) = P_x(k_x)/P_x(k_x = 0)]$ of conventional and precession LTEM images. We use the line profile within the red rectangles in the inset for each $\Delta f$. The gray-shaded region at low $k_y$ ($< 3.0$ μm$^{-1}$) is not used for analysis because it is strongly affected by a peak around $k_y = 0$. (c,d) $\widetilde{P}_y(k_y)$ of the conventional and precession LTEM images in the same region. (e,f) $C_{M+D} = \int dk_x \widetilde{P}_x(k_x)$ and $C_D = \int dk_y \widetilde{P}_y(k_y)$ for conventional and precession LTEM with the integration range from 3 μm$^{-1}$ to 10 μm$^{-1}$.